\documentclass[reprint,superscriptaddress,amssymb,amsmath,aps,showpacs,10pt,floatfix,prb,longbibliography]{revtex4-2}
\def\CVS{CsV$_3$Sb$_5$}
\def\KVS{KV$_3$Sb$_5$}
\def\RVS{RbV$_3$Sb$_5$}
\def\AVS{$A$V$_3$Sb$_5$}
\def\cm{cm$^{-1}$}
\def\Tc{$T_{\rm CDW}$}
\def\DCDW{$\Delta_{\rm CDW}$}
\def\wp{$\omega_p$}
\usepackage{graphicx}%
\usepackage{color}
\usepackage{epstopdf}
\usepackage{amssymb}
\usepackage{amsmath}
\usepackage{amsfonts}
\usepackage[note-name=, use-sort-key = false]{notes2bib}

\usepackage{xcolor}

\usepackage{color}
\usepackage[colorlinks,bookmarks=false,citecolor=darkblue,linkcolor=red,urlcolor=blue]{hyperref} 
\definecolor{darkred}{rgb}{0.7,0.0,0.0}

\definecolor{darkblue}{rgb}{0,0.02,0.45}

\definecolor{darkgreen}{rgb}{0.02,0.45,0.0}

\definecolor{violet}{rgb}{0.8,0.2,0.6}

\begin{document}
\title{Low-energy optical properties of the non-magnetic kagome metal \CVS}

\author{E. Uykur}
\email{ece.uykur@pi1.physik.uni-stuttgart.de}
\affiliation{1. Physikalisches Institut, Universit{\"a}t Stuttgart, 70569 
Stuttgart, Germany}

\author{B. R. Ortiz}
\affiliation{Materials Department and California Nanosystems Institute,
University of California Santa Barbara, Santa Barbara, CA, 93106, United States}

\author{O. Iakutkina}
\affiliation{1. Physikalisches Institut, Universit{\"a}t Stuttgart, 70569 
Stuttgart, Germany}

\author{M. Wenzel}
\affiliation{1. Physikalisches Institut, Universit{\"a}t Stuttgart, 70569 
Stuttgart, Germany}

\author{S. D. Wilson}
\affiliation{Materials Department, University of California Santa Barbara, Santa Barbara, CA, 93106, United States}

\author{M. Dressel}
\affiliation{1. Physikalisches Institut, Universit{\"a}t Stuttgart, 70569 
Stuttgart, Germany}

\author{A. A. Tsirlin}
\email{altsirlin@gmail.com}
\affiliation{Experimental Physics VI, Center for Electronic Correlations and Magnetism, University of Augsburg,
86159 Augsburg, Germany}
\date{\today}

\begin{abstract}
Temperature-dependent reflectivity measurements on the kagome metal \CVS\ 
in a broad frequency range of $50-20000$~cm$^{-1}$ down to $T$=10~K are reported. The charge-density wave (CDW) formed below \Tc\ = 94~K manifests itself in a prominent spectral-weight transfer from low to higher energy regions. The CDW gap of 75~meV is observed at the lowest temperature and shows significant deviations from an isotropic BCS-type mean-field behavior. Absorption peaks appear at frequencies as low as 200~cm$^{-1}$ and can be identified with interband transitions according to density-functional calculations. The change in the interband absorption compared to \KVS\ reflects the inversion of band saddle points between the K and Cs compounds. Additionally, a broader and strongly temperature-dependent absorption 
feature is observed below 1000~cm$^{-1}$ and assigned to a displaced Drude peak. It reflects localization effects on charge carriers.
\end{abstract}

\pacs{}
\maketitle

\section{Introduction}

Kagome lattices are the topic of extensive research as they provide a suitable playground for the realization of exotic phases \cite{Liu2019}. Recent efforts are not limited to the well-known insulating quantum spin liquid candidates \cite{Broholm2020}, but extend to the newer members of the class, the so-called kagome metals, where the geometrical frustration meets with itinerant charge carriers. The peculiar geometry of the kagome framework leads to the coexistence of flat bands and linearly dispersing topological Dirac bands that have been confirmed in several magnetic kagome metals \cite{Ye2018,  Lin2018, Yin2019, Kang2020}.  

The discovery of the \AVS\ ($A$ = K, Cs, Rb) compounds has started a new direction in the study of kagome metals \cite{Ortiz2019}. \AVS\ crystallize in the $P6/mmm$ space group, where the V-Sb kagome networks are separated with Sb graphite-like layers and the $A$ alkali ions, creating a well-isolated situation for the kagome layers. The resultant system can be identified as quasi-2D and displays several interesting properties. A clear transport and magnetic anomaly is reported at \Tc\ = 94~K for \CVS\ \cite{Ortiz2020} (78~K for \KVS\ \cite{Ortiz2019} and 102~K for \RVS\ \cite{Yin2021}) and tied to a charge-density wave (CDW) formation. Furthermore, superconductivity develops below $T_{\rm  c}$=2.5~K for \CVS\ \cite{Ortiz2020} (below 1~K for \KVS\ \cite{Ortiz2021} and \RVS\ \cite{Yin2021}). The coexistence of the density-wave order with superconductivity attracted immediate attention, and the tunability of both orders via external pressure has been demonstrated by transport measurements \cite{Zhao2021, Zhang2021, Du2021,Chen2021, Chen2021a}. 

While the energy scales and the order parameters of both CDW and superconducting phases remain a source of considerable debate, several recent experimental observations identify \CVS\ and its siblings as highly unconventional materials. Large anomalous Hall effect in the absence of magnetic ordering \cite{Yang2020, Kenney2020,Yu2021}, anisotropic and possibly multiple gaps in the density-wave and superconducting states \cite{Wang2021, Nakayama2021, Zhao2021, Xu2021, Xiang2021, Ni2021, Duan2021, Chen2021b}, and re-entrance of superconductivity at elevated pressures \cite{Chen2021, Zhang2021, Lin2021, Denner2021} all suggest peculiarities of the underlying band structure. Some of the experimentally observed phenomena are indeed anticipated for the kagome metals \cite{Wu2021, Tan2021, Denner2021, Lin2021, Park2021}. This renders the \AVS\ compounds convenient model systems that can realize the peculiar kagome physics in the real-world environment. 

Optical spectroscopy is a powerful tool for studying the electronic structure and especially its changes caused by density-wave instabilities. Here, we examine the optical properties of \CVS\ in a broad energy and temperature range at temperatures both above and below \Tc. We observe several interband transitions that distinguish \CVS\ from its \KVS\ sibling~\cite{Uykur2021} despite a close similarity between the band structures of the two compounds. Even more intriguingly, the conventional Drude response of free charge carriers is accompanied by a prominent, temperature--dependent localization peak indicative of the localized charge carriers. This peak is a fingerprint of interactions in \CVS\ beyond the simple band scenario.

The interband transitions have been investigated with the help of band structure and band-resolved optical conductivity calculations. While the low-energy interband transitions are accumulated to a sharp peak for the \KVS\ compound, they are spread over a broader energy range in \CVS\ and extend to lower energies. These
changes can be attributed to modifications of the band structure in the vicinity of the $M$ point, the region deemed crucial for electronic instabilities of a kagome metal~\cite{Kiesel2013, Park2021, Wu2021}.

\section{Methods}
\subsection{Experimental}

High-quality single crystals were synthesized as explained elsewhere \cite{Ortiz2019, Ortiz2021}. A sample with the lateral dimensions of $\sim 5\times5$~mm$^2$ and thickness of about 100 $\mu$m was used for the optical measurements. This sample was cleaved prior to the experiment. The CDW transition temperature and the possible Cs-deficiency of our sample were monitored via four-point resistivity measurements performed on the same piece as used in optics. The obtained resistivity agrees with the literature data~\cite{Ortiz2020} and confirms the \CVS\ stoichiometry. The kink marks the density-wave transition at 94~K, also in agreement with the previous literature~\cite{Ortiz2020}. 

Temperature- and frequency-dependent reflectivity measurements have been performed utilizing two different Bruker Fourier transform infrared (FTIR) spectrometers down to 10~K and between 5 meV-2.5 eV (50-20000~\cm). A Vertex80v spectrometer coupled with an Hyperion IR microscope is used for the high-energy range ($\omega>600$~ \cm), while an IFS113v spectrometer with a custom-built cryostat was used for the low-energy measurements. Freshly evaporated gold mirrors served as the reference in the measurements. The absolute value of the reflectivity is deduced with the gold-overcoating technique in the far-infrared range. 

Optical conductivity is calculated from the measured in-plane reflectivity spectra via the Kramers-Kronig analysis. We utilized standard Hagen-Rubens extrapolation below 50~\cm\ and x-ray scattering functions for the high-energy extrapolations \cite{Tanner2015}. 

The skin depth of the sample reaches around 1000 nm at the far-infrared range and exceeds the penetration depth of the probing radiation for all the measured temperatures and frequencies indicating that the current measurements reflect the bulk nature of the probed sample. 

\begin{figure*}
	\centering
	\includegraphics[width=2\columnwidth]{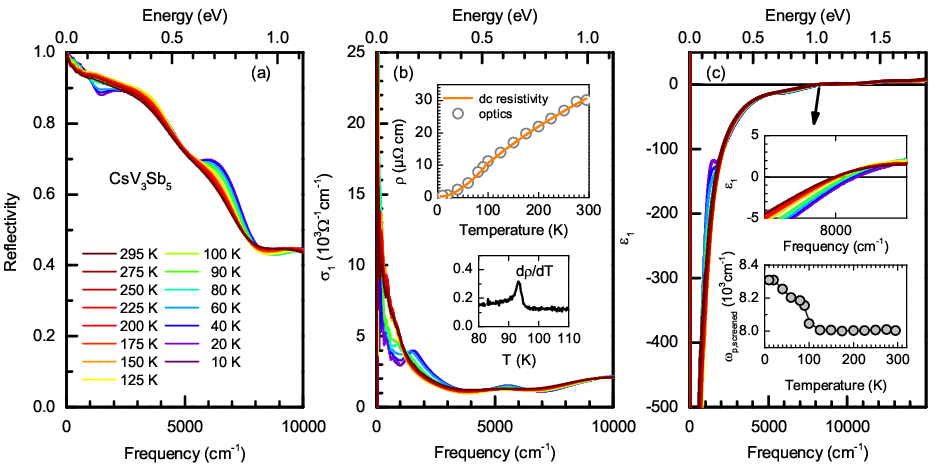}
	\caption{(a) Temperature-dependent reflectivity in the $ab$-plane. (b) Calculated optical conductivity in a broad frequency and temperature range. Inset: dc-resistivity measured on the same crystal as used in the optical measurements, overlaid with the values obtained from the Hagen-Rubens fit of the reflectivity. First derivative of the dc-resistivity curve marks the CDW transition at 94~K. (c) Temperature- and frequency-dependent 
dielectric permittivity. Zero crossing, the screened plasma frequency, is 
marked in the inset and shows a clear blue shift below \Tc. Temperature dependence of the plasma frequency is also shown demonstrating the clear increase across the CDW transition. }		
	\label{Ref_OC}
\end{figure*}

\subsection{Computational}

Density-functional (DFT) band-structure calculations and calculations of the optical conductivity were performed in the \texttt{Wien2K}~\cite{wien2k,Blaha2020} code using Perdew-Burke-Ernzerhof flavor of the exchange-correlation potential~\cite{pbe96}. Experimental structural parameters from Ref.~\cite{Ortiz2020} were chosen for the undistorted \CVS{} structure, whereas possible CDW structures were obtained by a structural relaxation in \texttt{VASP}~\cite{vasp1,vasp2} similar to Ref.~\cite{Uykur2021}. Spin-orbit coupling was included for the calculations of band structure and optical conductivity. Self-consistent calculations and structural relaxations were converged on the $24\times 24\times 12$ $k$-mesh for the undistorted \CVS\ structure (normal state) and $12\times 12\times 12$ $k$-mesh for the distorted structures (CDW). Optical conductivity was calculated on 
the $k$-mesh with up to $100\times 100\times 50$ points for the normal state and $36\times 36\times 36$ points for the CDW state. 

\section{Results and Discussion}

\subsection{Optical spectra}

The temperature-dependent reflectivity and optical conductivity of \CVS\ are given in Fig.~\ref{Ref_OC} (a) and (b), respectively, over a broad frequency range. At high temperatures, the increase in the reflectivity toward lower frequencies demonstrates metallic nature of the sample. The onset of the reflectivity around 8000~\cm\ marks the plasma frequency. Consistently, optical conductivity shows the Drude-like increase at lower frequencies followed by an almost flat behavior above 3000~cm$^{-1}$. The details of the low-energy regime will be discussed in more detail in the later sections. 

Across the density-wave transition, prominent changes are seen in both reflectivity and optical conductivity. Reflectivity develops a dip around 0.15~eV, mirrored by the spectral weight (SW) transfer in the optical conductivity and the peak developing below \Tc. The conductivity values in the $\omega\rightarrow 0$ limit, as obtained from the Hagen-Rubens fits of the reflectivity, coincide with the four-probe dc resistivity measurements performed on the same sample as shown in the inset of Fig.~\ref{Ref_OC}(b).

The free carrier dynamics in \CVS\ also change across the density-wave transition. The screened plasma frequency of the compound can be estimated via zero-crossing of the dielectric permittivity ($\epsilon_1$) \cite{Dressel2002} as given in Fig.~\ref{Ref_OC}(c). The overall behavior reflects the highly metallic nature of the sample, as $\epsilon_1$ is negative and diverges at $\omega\rightarrow 0$. While a small absorption feature is also visible in $\epsilon_1$ below \Tc, the behavior in this energy range is governed by the intraband transitions. Although the true plasma frequency is masked by the interband absorptions at higher energies, it can be estimated as $\omega_p^{\rm screened} = \omega_p / \sqrt{\epsilon_\infty}$, where $\epsilon_\infty$ is mostly determined by the high-energy contributions and stays around 10 at all temperatures. The plasma frequency, on the other hand, starts increasing below \Tc\ as demonstrated in the insets of Fig.~\ref{Ref_OC}(c). 

Plasma frequency in the optical measurements defines the ratio $\omega_p = \sqrt{4\pi ne^2/m^*}$ between the free carrier density, $n$, and the effective mass, $m^*$. Therefore, the changes in \wp\ cannot be interpreted in a straight-forward manner. Recent Hall effect measurements revealed the non-monotonic change in the carrier concentration with temperature \cite{Yu2021}. A small decrease followed by an increase in the carrier concentration below \Tc\ has also been shown in the case of \KVS\ \cite{Yang2020}. The increase in \wp\ below \Tc\, hence, can be attributed to the increasing carrier density in \CVS. The unchanged plasma frequency above \Tc, on the other hand, suggests that the effective mass should be renormalized in this temperature range and counteracts the decrease in the carrier density. We note in passing that such a strong change in \wp\ below \Tc\ has not been observed in \KVS~\cite{Uykur2021}, suggesting a more pronounced increase in the effective mass across \Tc\ in \CVS. 

\subsection{Energy scale of the charge-density wave}

\begin{figure} [h]
	\centering
	\includegraphics[width=1\columnwidth]{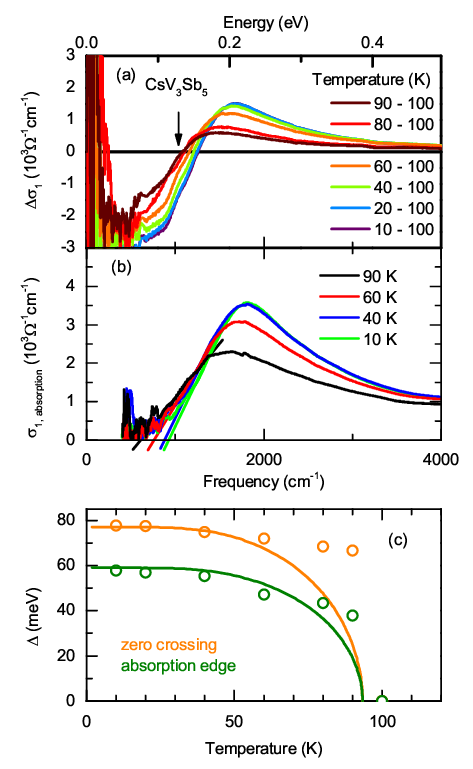}
	\caption{(a) Difference optical conductivity ($\Delta\sigma_1=\sigma_1(T) - \sigma_1(100~K)$) spectra in the CDW region. The 100~K data measured right above the transition are chosen as the base curve. The zero crossing in the difference curves marks the CDW gap (note that the optical measurements detect the transitions across the Fermi energy and hence the observed value corresponds to $2\Delta$) (b) Optical conductivity at the absorption edge. The extrapolation of the steepest part to the frequency axis is taken as the $2\Delta$ (c) Temperature dependence of the gap obtained from the zero crossing and the extrapolation of the absorption edge clearly deviates from the BCS-type mean-field behavior (solid lines) above 50~K. }		
	\label{CDW}
\end{figure}

The difference optical conductivity clearly demonstrates the density-wave like spectral weight redistribution as given in Fig.~\ref{CDW}. The SW suppressed at low energies is recovered in the higher-energy range around 0.5~eV with a maximum at 0.2~eV. Here, we utilized the zero-crossing point as the scale for the density-wave gap (2$\Delta$) in line with the previous studies~\cite{Schafgans2012}. Alternatively, we subtracted the Drude part at each temperature and extrapolated the steepest part of the edge to zero with a straight line. This procedure led to the very similar values of \DCDW\ and their temperature dependence. We also note that positive values seen in the difference spectra at very low energies in Fig.~\ref{CDW}(a) are caused by the sharpening of the Drude peak and do not affect the estimate of the gap using zero crossing. Moreover, our \DCDW\ value at low temperatures is consistent with the recent estimate from ARPES on \CVS\ \cite{Wang2021}.

In principle, the thermodynamics of the CDW state should resemble that of a superconducting state \cite{Gruener1988}. Therefore, it is natural to compare the temperature dependence of the CDW gap with the characteristic BCS-type form, where the gap vanishes at \DCDW\ $= 1.76k_{\rm B}$\Tc. This simple assumption yields a CDW transition temperature of around 513~K for the obtained gap value of $\sim$~78~meV and significantly exceeds the experimental value of 94~K. The mean-field description of CDW formation neglects the fluctuation phenomena, which might be at the center of this mismatch and especially become important in higher dimensions (d$>$1). The general picture reveals that the fluctuations should strongly suppress the CDW transition temperature below the mean-field estimate (\Tc\ $< T_{\rm mean-field}$). Furthermore, even though directional correlations can form below \Tc, the long range order may only be formed when the neighboring correlations start to couple together leading to a temperature scale of $T_{\rm long-range} <$ \Tc\ $<T_{\rm mean-field}$. Considering the 2D nature of the studied system, it may also be relevant in the current case. 

It is further remarkable that not only \DCDW/\Tc\ at the lowest temperature, but also the temperature dependence of \DCDW\ deviates from the BCS behavior. Fig.~\ref{CDW}(b) shows that the gap opening in \CVS\ is more of a first-order transition type, with an abrupt increase right below the transition temperature, as also observed in the \KVS\ sister compound~\cite{Uykur2021}. The energy scale in \CVS\ is slightly elevated compared to \KVS, in agreement with the higher \Tc\ of the Cs compound. Some of the recent reports suggested multiple energy scales of the CDW gap in \CVS\ according to photoemission measurements~\cite{Nakayama2021,Wang2021}. While our present data can not resolve these subtle details, the deviations from the mean-field temperature evolution of $\Delta$ corroborate previous observations, because in the presence of a strongly anisotropic gap or multiple gaps the assumption of the isotropic CDW gap does not hold. It is worth noting that a similar non-BCS evolution of the gap has been seen in \KVS~\cite{Uykur2021}, indicating that the CDW gap in the K compound may be anisotropic, as well.

\subsection{Localization peak}

\begin{figure}
	\centering
	\includegraphics[width=1\columnwidth]{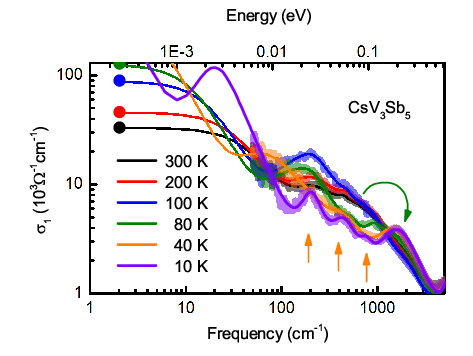}
	\caption{Low energy optical conductivity for selected temperatures. The thin-solid lines are Drude-Lorentz fits to the optical conductivity as described in the text. Please note that for 10~K spectrum, the localization peak cannot be observed but estimated based on the linear shift with respect to the peaks observed at higher temperatures. The solid circles are the dc conductivity values. Orange arrows show the low-energy absorption bands, while the green arrow demonstrates the spectral weight transfer to a peak below \Tc. }	
	\label{OC_LE}
\end{figure}

\begin{figure}
	\centering
	\includegraphics[width=1\columnwidth]{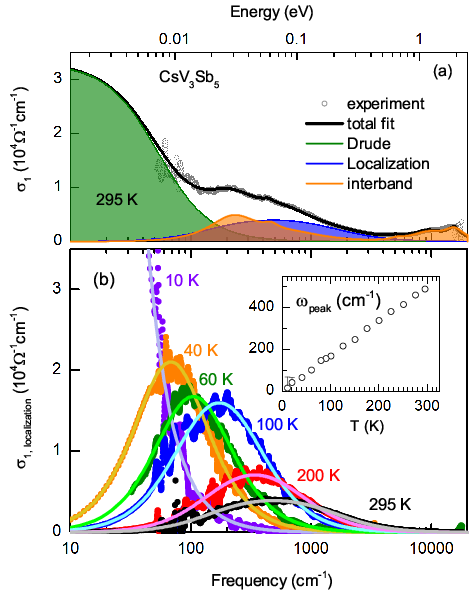}
	\caption{(a) A sample decomposition of the optical conductivity at room temperature consists of a Drude peak (green), a localization peak (blue), 
and multiple peaks due to interband transitions (orange). (b) The temperature dependence of the localization peak obtained by subtracting the Drude and interband contributions from the experimental optical conductivity. The fits to the localization peak are also shown as a guide to the eye. Inset: Temperature evolution of the peak position.}		
	\label{components}
\end{figure}

Having described the general trends in the optical conductivity, we now turn to decomposition of the experimental spectra. The low-energy optical conductivity is demonstrated in Fig.~\ref{OC_LE} for selected temperatures above and below the CDW transition, where several contributions to the spectra are visible. Different contributions were modeled with the phenomenological Drude-Lorentz approach and also presented in Fig.~\ref{components}(a) along with the experimental spectra.

\begin{align}
\label{Eps}
&\tilde{\varepsilon}=\varepsilon_1 + i\varepsilon_2,\\\notag
&\tilde{\varepsilon}(\omega)= \varepsilon_\infty - \frac{\omega^2_{p,{\rm Drude}}}{\omega^2 + i\omega/\tau_{\rm\, Drude}} + \sum\limits_j\frac{\Omega_j^2}{\omega_{0,j}^2 - \omega^2-i\omega\gamma_j}.
\end{align}
Here $\varepsilon_\infty$ stands for high-energy contributions to the real part of the dielectric permittivity. $\omega_{p,{\rm Drude}}$ and $1/\tau_{\rm\,Drude}$ are the plasma frequency and the scattering rate of the itinerant carriers, respectively. $\omega_{0,j}$, $\Omega_j$, and $\gamma_j$ describe the resonance frequency, width, and the strength of the $j^{th}$ excitation. The complex optical conductivity is then obtained as
\begin{align}
&\tilde{\sigma}=\sigma_1 + i\sigma_2,\notag\\
&\tilde{\sigma}(\omega)= -i\omega[\tilde{\varepsilon}-\varepsilon_\infty]/4\pi.
\label{Cond}
\end{align}

As an example, a decomposition of the spectra is given in Fig.~\ref{components}(a) for the room-temperature data. The low-energy upturn of the conductivity is described with the Drude contribution, where the dc-conductivity is fixed to the value obtained from the Hagen-Rubens fit. We used several Lorentzians for the interband transitions, as demonstrated by the orange curve in the sample fit of Fig.~\ref{components}(a). Three of these Lorentzians are placed at fairly low energies, as shown by the arrows in Fig.~\ref{OC_LE}. They do not change significantly with temperature except getting sharper. Another absorption feature in the same low-energy range (blue curve) is, on the contrary, strongly temperature-dependent and  assigned to a so-called localization peak, in line with earlier optical studies of the kagome metals~\cite{Biswas2020, Uykur2021}. This absorption is clearly distinguished from the interband transitions by its strong red-shift upon cooling. An additional argument for the distinct nature of this peak comes from DFT calculations (see below) that perfectly reproduce the interband absorption but not the additional peak, which should then arise from an intraband process.

\begin{figure}
	\centering
	\includegraphics[width=1\columnwidth]{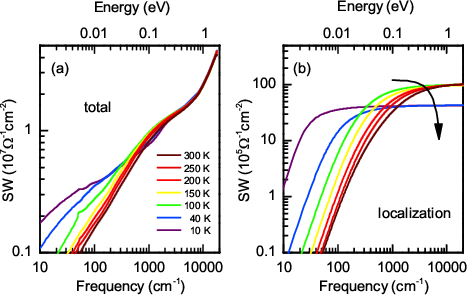}
	\caption{(a) Temperature-dependent total spectral weight. The overall SW 
is conserved within the measured energy range. (b) Temperature dependence 
of the localization peak SW obtained by the integration of the Lorentzian 
fits to the experimental curves shown in Fig.~\ref{components}(b). An abrupt decrease in the SW across \Tc\ is visible.}		
	\label{SW}
\end{figure}

The temperature evolution of this localization peak is shown in detail in 
Fig.~\ref{components}(b) along with the temperature dependence of the peak position in the inset. After the overall fit of the spectra, we subtracted both Drude and interband contributions from the experimental curves to highlight the behavior of the localization peak. With decreasing temperature, the localization peak gradually shifts toward lower energies and eventually moves out of our measurement range. Our detailed analysis [Fig.~\ref{SW}] shows that the overall SW is conserved within the measured energy range in accord with the optical sum rules. On the other hand, the SW of the localization peak decreasing across the transition suggests a partial spectral weight transfer from the localization peak to the high-energy absorption as the CDW order develops (Fig.~\ref{OC_LE}). This observation strengthens the assumption that the localization peak should be understood as arising from an intraband process, which is affected by the partial gap opening and consequent reduction in the density of states at the Fermi level.

A low-energy absorption feature with a distinct temperature dependence is a common occurrence in the world of strongly correlated electronic systems. Its microscopic origin, on the other hand, remains a source of considerable debate. One possibility is that dynamical localization effects take place, caused by the interaction of charge carriers with low-energy degrees of freedom, such as phonons, magnetic or charge fluctuations. This leads to backscattering of electrons and a displaced Drude peak, with the zero-frequency response of the charge carriers shifted to a finite frequency value~\cite{Pustogow2021}. Our aforementioned analysis is based on this displaced Drude formalism, as presented in Ref.~\cite{Fratini2014}. The common occurrence of the displaced Drude peak in strongly correlated electronic systems~\cite{Luca2017} indicates the possible importance of electron-electron interactions in \CVS.

\begin{figure*}
	\centering
	\includegraphics[width=2\columnwidth]{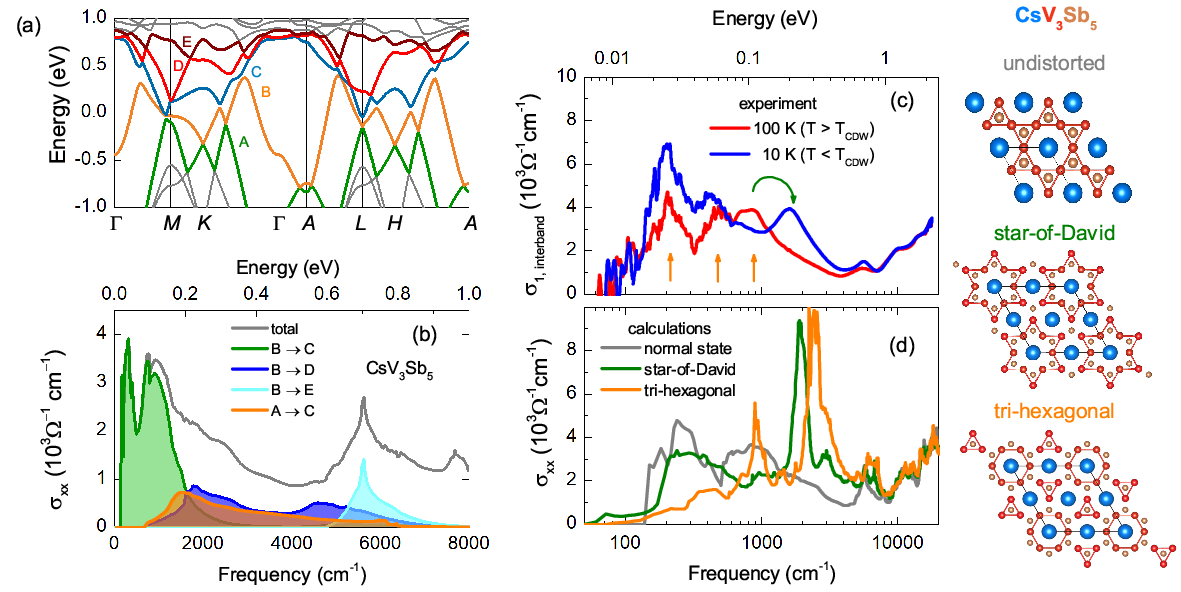}
    \caption{(a) Band structure of \CVS. (b) Band-resolved optical conductivity. (c) Experimental interband transitions at 100~K (normal state) and at 10~K (CDW state). Orange and green arrows mark the peak positions of the interband transitions and the spectral weight transfer to a peak at higher energy range below \Tc. (d) Calculated optical conductivity in the normal and CDW states given for two types of distortions (star-of-David and tri-hexagonal). Right panel shows the undistorted and the distorted structures.}			
	\label{BR}
\end{figure*}

\subsection{Band-resolved optical conductivity}

Further insight into the low-lying interband transitions is obtained from 
DFT calculations of the band structure and optical conductivity, Fig.~\ref{BR}(a) and (b). In accord with the in-plane measurement configuration, we consider $\sigma_{xx}$ component of the calculated optical conductivity. In the normal state, the low-energy absorption is almost solely related to the transitions between bands $B$ and $C$ in the vicinity of the $L$ 
and $M$ points of the Brillouin zone. The higher-energy absorptions due to the $A\rightarrow B$ and $B\rightarrow D$ transitions are relatively frequency-independent, as expected for the transitions between linearly dispersing bands. The pronounced mid-infrared absorption around 0.7~eV mainly results from the transitions between bands $B$ and $E$. Unlike bands $A-D$, band $E$ is weakly dispersive and can be understood as a flat band of the kagome metal. Therefore, the absorption peak at 0.7~eV most likely reflects the presence of a flat band pinned at this energy.

In Fig.~\ref{BR}(c) and (d), we show a comparison between the experimental interband transitions and DFT results for both normal and CDW states of \CVS. For the CDW state, we consider star-of-David and tri-hexagonal (inverse star of David) models as two possible distortions of the vanadium kagome planes~\cite{Uykur2021,Tan2021,Ortiz2021a,Ratcliff2021}, see the right part of Fig.~\ref{BR}. Both models successfully describe the shift of the spectral weight toward higher energies and the absorption peak placed at higher energies compared to the normal state. The peak position is better reproduced by the star-of-David CDW model that further reveals a dip in the optical conductivity at $700-1500$~cm$^{-1}$ and a low-energy peak at $200-500$~cm$^{-1}$, in qualitative agreement with the experiment and at odds with the calculations for the tri-hexagonal CDW structure. This result is somewhat unexpected, because the tri-hexagonal CDW is more stable and lies 12.0~meV/f.u. lower in energy than the undistorted structure, compared to the stabilization energy of only 3.2~meV/f.u. for the star-of-David CDW. These stabilization energies are in line with the concurrent DFT studies of \CVS~\cite{Tan2021,Ortiz2021a,Ratcliff2021}.

\begin{figure}
	\centering
	\includegraphics[width=1\columnwidth]{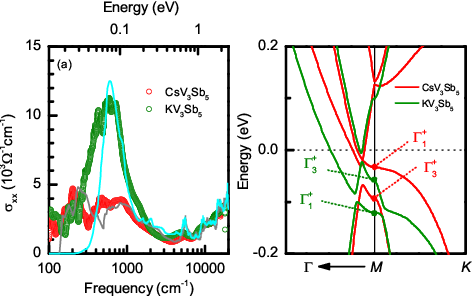}
	\caption{ (a) Normal-state interband transitions in \CVS\ and \KVS, as seen experimentally (symbols) and calculated with DFT (lines). Experimental data are taken at 100~K for \CVS\ and 90~K for \KVS, right above \Tc. (b) A comparison of the \CVS\ and \KVS\ band structures in the $k_z=0$ plane. Note 
the difference around $M$ in the vicinity of the saddle points. }		
	\label{comparison}
\end{figure}

It is also instructive to compare low-energy spectral features of the two 
sister compounds \CVS\ and \KVS. At first glance, their band structures and transport properties are very similar. Optical studies reveal several similarities too, namely, the presence of the low-lying interband transitions, the localization peak, and the highly metallic character reflected in the sharp Drude peak. Some striking differences are not to be overlooked, though. (i) The carrier scattering in \CVS\ is stronger than in \KVS, resulting in a broader Drude peak that is well seen within the energy range of our optical measurement and partially overlaps with the localization and interband peaks. (ii) While the high-energy interband transitions in the two compounds are almost identical, the lower-energy part of the spectrum changes significantly and evolves from one intense peak in \KVS\ to a characteristic structure with three weaker peaks below 1000~cm$^{-1}$ in \CVS\ [Fig.~\ref{comparison}(a)]. These changes are well reproduced by DFT (albeit with a shift of the Fermi energy required for \KVS~\cite{Uykur2021}) and reflect an important change in the band dispersions around $M$ where saddle points occur right below the Fermi level [Fig.~\ref{comparison}(b)]. 

These saddle points are a direct consequence of the underlying kagome structure and correspond to the van Hove singularity of a simple kagome metal~\cite{Kiesel2013}. In the multi-orbital case of \AVS, several saddle points appear~\cite{Denner2021}. Using the notation of Ref.~\cite{Park2021}, we identify $\Gamma_1^+$ at $-30$~meV and $\Gamma_3^+$ at $-95$~meV in \CVS, to be compared with, respectively, $-120$~meV and $-60$~meV in \KVS. This inversion of the saddle points affects band dispersions along $M-\Gamma$. Two bands cross very close to the Fermi level in \CVS, while in \KVS\ the band crossing is avoided because bands of different nature meet around this point. The presence of the band crossing at $-20$~meV, in the immediate vicinity of the Fermi level, results in the interband transitions extending to lower energies in \CVS\ compared to \KVS.

\section{Conclusions}

In conclusion, we investigated optical properties of the kagome metal \CVS\ in a broad frequency and temperature range. The CDW transition manifests itself by the spectral weight transfer toward higher energies, indicating a gap opening and the reduction in the density of states at the Fermi level. Temperature dependence of the gap indicates an abrupt, first-order transition and strongly deviates from the BCS-type mean-field behavior. The obtained energy of the CDW gap is slightly higher compared to \KVS, in line with the slightly higher \Tc\ of the Cs compound. 

The low-energy optical conductivity is dominated by interband transitions along with a prominent localization peak that gradually shifts to lower energies upon cooling. Across the CDW transition, a decrease in the spectral weight of this localization peak is observed, indicating the involvement of the localized carriers in the CDW formation. The presence and persistence of this localization peak may hint toward the correlated nature of \CVS.

Interband transitions identified via DFT calculations show a distinct three-peak structure at low energies, which is manifestly different from the single interband absorption peak in \KVS. We ascribe this difference to the inversion of band saddle points that are presently discussed as the possible driving force of the CDW transition and other peculiarities of the \AVS\ physics~\cite{Park2021,Wu2021,Denner2021}.\\

\textbf{Note added:} During the preparation of this manuscript, another optical
study on CsV$_3$Sb$_5$ appeared \cite{Zhou2021}. The reported \Tc\ is slightly lower and the dc resistivity is slightly higher than in our study, indicating a possible difference in the stoichiometry of the sample. Our optical data qualitatively agree with Ref.~\cite{Zhou2021}, but the interpretation is remarkably different. Zhou $et$ $al.$ \cite{Zhou2021} analyzed their spectra using the two-Drude formalism, with both Drude peaks centered at zero frequency, unlike in our study where one of the peaks is displaced and becomes a localization peak. At 200 K, these two Drude peaks of Ref.~\cite{Zhou2021} contain most of the spectral weight below 4000 cm$^{-1}$, at odds with our DFT results, band-resolved optical conductivity, and the very
nature of the CsV$_3$Sb$_5$ band structure that allows multiple interband transitions in this frequency range.

\begin{acknowledgments}
Authors acknowledge the fruitful discussion with Simone Fratini and  technical support by Gabriele Untereiner. S.D.W. and B.R.O. gratefully acknowledge support via the UC Santa Barbara NSF Quantum Foundry funded via the Q-AMASE-i program under award DMR-1906325. B. R. O. also acknowledges support from the California NanoSystems Institute through the Elings fellowship program. The work has been supported by the Deutsche Forschungsgemeinschaft (DFG) via DR228/51-1 and UY63/2-1. E.U. acknowledges the European Social Fund and the Baden-W\"urttemberg Stiftung for the financial support of this research project by the Eliteprogramme. A.T. was supported by the Federal Ministry for Education and Research via the Sofja Kovalevskaya Award of Alexander von Humboldt Foundation.
\end{acknowledgments}

\bibliography{CsV3Sb5}

\end{document}